\pdfoutput=1

\documentclass[journal]{IEEEtran}
\usepackage[pdftex]{graphicx}

\ifCLASSINFOpdf
\else
\fi
\usepackage{amsmath}

\usepackage{multirow}

\usepackage[]{hyperref}
\hypersetup{
    pdftitle={Scalability analysis of large-scale LoRaWAN networks in ns-3},
    pdfauthor={Floris Van den Abeele},
    pdfsubject={Scalability analysis of large-scale LoRaWAN networks in ns-3},
    bookmarksnumbered=true,     
    bookmarksopen=true,         
    bookmarksopenlevel=1,       
    colorlinks=true,
    allcolors=blue,
    pdfstartview=Fit,           
    pdfpagemode=UseOutlines,
    pdfpagelayout=TwoPageRight
}
\usepackage{hypcap}

\begin{document}
%
\title{Scalability analysis of large-scale LoRaWAN networks in ns-3}
%
%
%

\author{Floris Van den Abeele, Jetmir Haxhibeqiri, Ingrid Moerman, Jeroen Hoebeke\\
Ghent University -- imec, IDLab, Department of Information Technology\\
Technologiepark Zwijnaarde 15, B-9052 Ghent, Belgium\\
\{floris.vandenabeele,jetmir.haxhibeqiri,ingrid.moerman,jeroen.hoebeke\}@ugent.be}

\markboth{Journal of \LaTeX\ Class Files,~Vol.~14, No.~8, August~2015}%
{Shell \MakeLowercase{\textit{et al.}}: Bare Demo of IEEEtran.cls for IEEE Journals}
%



\maketitle

\begin{abstract}
As LoRaWAN networks are actively being deployed in the field, it is important to comprehend the limitations of this Low Power Wide Area Network technology.
Previous work has raised questions in terms of the scalability and capacity of LoRaWAN networks as the number of end devices grows to hundreds or thousands per gateway.
Some works have modeled LoRaWAN networks as pure ALOHA networks, which fails to capture important characteristics such as the capture effect and the effects of interference.
Other works provide a more comprehensive model by relying on empirical and stochastic techniques.
This work uses a different approach where a LoRa error model is constructed from extensive complex baseband bit error rate simulations and used as an interference model.
The error model is combined with the LoRaWAN MAC protocol in an ns-3 module that enables to study multi channel, multi spreading factor, multi gateway, bi-directional LoRaWAN networks with thousands of end devices.
Using the lorawan ns-3 module, a scalability analysis of LoRaWAN shows the detrimental impact of downstream traffic on the delivery ratio of confirmed upstream traffic.
The analysis shows that increasing gateway density can ameliorate but not eliminate this effect, as stringent duty cycle requirements for gateways continue to limit downstream opportunities.
\end{abstract}


%
\IEEEpeerreviewmaketitle

\section{Introduction}
With the ongoing continuous growth of the Internet of Things, the number of IoT application domains and deployments continues to increase.
Market forecasts illustrating this growth, estimate that the number of connected IoT devices will continue to grow at an annual rate of 32\% and will reach 20.8 billion IoT end points by the end of this decade~\cite{Gartner2015}.
Some of these novel IoT applications require low-rate, long-range and delay-tolerant wireless communication at very low energy usage and cost. 
These types of requirements are hard to fulfill using traditional Machine to Machine technologies such as cellular or WPAN~\cite{LpwanoverviewRaza2017}.
Low Power Wide Area Networks~(LPWANs) are a new set of technologies that are designed to fill this gap in traditional technologies.
By combining low energy usage with long range communication, they promise to bring connectivity that suits large scale, low power and low cost IoT deployments with battery lives up to ten years~\cite{LpwanoverviewRaza2017}. 
The 2016 Cisco VNI 2015-2020 data traffic forecast estimates that the share of LPWA networks in global M2M connections will grow from 4\% in 2015 to 28\% in 2020~\cite{cisco-vni-2016}.
Similar market share numbers for LPWANs are reported in~\cite{nokia-iot-connectivity}.

LoRaWAN~\cite{alliance2015lorawan} is an LPWAN technology that builds on top of the LoRa modulation scheme, which is developed by Semtech.
The LoRaWAN alliance has standardized LoRa radio usage in sub-GHz unlicensed spectrum for most areas in the world.
By combining sub-GHz propagation and the LoRa modulation, LoRaWAN networks can cover large areas with only limited amounts of infrastructure.
LoRaWAN networks are being deployed today. 
For instance in Belgium, Proximus, a large telecommunications company, provides LoRaWAN coverage in the whole of Flanders and in the major cities in Wallonia~\cite{ProximusNV}. 
Another interesting initiative is The Things Network, where a community of mostly volunteers is collaborating to build a world-wide LoRaWAN network.

While LoRaWAN networks are already being deployed in the field, a number of questions remain unanswered about their performance as said networks grow larger and larger.
Most LPWAN technologies promise to connect a massive number of devices~(e.g. tens of thousands of devices per LoRaWAN gateway), but how valid is this claim in the case of LoRaWAN networks?
What is the impact of network parameters on large-scale LoRaWAN networks?
Apart from a large number of devices, what is the impact of multiple gateways in large-scale networks?
Does increasing gateway density yield measurable network performance benefits?

The work presented here aims to provide answers to the research questions posed above by studying LoRaWAN networks in the ns-3 network simulator.
A network simulator provides the flexibility to relatively quickly study a large number of different LoRaWAN scenarios at the expense of accuracy due to limitations in the modeling complexity.
As such, the presented ns-3 module allows to study LoRaWAN networks with a varying numbers of end devices and gateways, different traffic types and patterns, different data rates, different (re)transmission and receive parameters and many other parameters.
Finally, the presented work also allows to study the impact of more fundamental changes to LoRaWAN network mechanics as everything is implemented in software. 

While a number of existing works have studied the scalability of LoRaWAN networks, they do not take into account the impact of downstream traffic, interference between transmitters and the presence of multiple LoRaWAN gateways in dense deployments.
Additionally, the spectrum modeling technique newly introduced in ns-3.26, which is applied in the LoRaWAN ns-3 module, should enable the module to be used in future coexistence studies with heterogeneous technologies such as 802.11ah.
A more detailed comparison with works from literature is available in section~\ref{sec:related-work}.

The contributions of this work are as follows.
Firstly, we built an error model for the LoRa modulation for different coding rates and spreading factors.
Secondly, we developed a comprehensive implementation of the LoRaWAN standard in the ns-3 simulator with support for class A end devices, multi gateway networks and an elementary network server.
Thirdly, we conducted a scalability study focusing on the impact of confirmed versus unconfirmed messages and the impact of downstream traffic in large-scale LoRaWAN networks.

Before detailing how LoRaWAN networks are modeled in ns-3 in section~\ref{sec:lorawan-ns3-module}, the next section provides the necessary background on LoRa and LoRaWAN in order to comprehend the modeling efforts in the ns-3 module.
Section~\ref{sec:scalability-analysis} presents the scalability analysis itself.
The results of the analysis are discussed and are compared to literature in section~\ref{sec:related-work}.

\section{Background: LoRa, LoRaWAN and ns-3}
As a Low Power Wide Area Network technology, LoRaWAN networks offer benefits such as large coverage areas and long battery life operation for end devices.
Unlike conventional network technologies~(e.g. cellular and LAN), the trade-off in data rate versus range leans heavily towards range in LoRaWAN.
LoRaWAN networks further employ a proprietary physical modulation technique~(named LoRa) which has been developed with long range and low power operation at its core.
This section describes some of the aspects of LoRaWAN networks that are important for the remainder of this paper.

The LoRa physical modulation is a chirp spread spectrum~(CSS) technique where the base symbol is an up-chirp.
An up-chirp is a signal in which the frequency increases with time. 
More specifically, in LoRa the frequency increases linearly with time so that the frequency of the up-chirp sweeps the entire bandwidth of the signal.
The constellation diagram of LoRa consists of time-shifted up-chirps.
Therefor, at the receiver the demodulation process attempts to determine the time shift in the received up-chirp.

Next to bandwidth, the spreading factor is a second important parameter of the LoRa modulation as it provides the flexibility of trading range for data rate.
The spreading factor can range from seven to twelve and determines the LoRa symbol rate as: $\mathrm{R_s = \frac{BW}{2^{SF}}}$.
As the spreading factor increases the symbol rate is lowered, thereby trading reduced data rate for increased range.
For a bandwidth of 125kHz, the PHY data rates range from 6835bps to 365bps for SF seven to twelve.
The different spreading factors are orthogonal. 
This means that a LoRa gateway can receive multiple transmissions on different spreading factors simultaneously.
Note that SF bits are mapped per LoRa symbol.

In order to further increase the robustness of the LoRa modulation, additional techniques such as forward error correction and interleaving are employed. 
These techniques are discussed in section~\ref{sec:lora-phy-baseband-implementation}.
Additionally, LoRa radios operate in the sub-GHz unlicensed bands as they provide a good trade-off between available unlicensed spectrum and reduced path loss.
The combination of these design considerations results in a high link budget at the expense of data rate, which means that LoRa transmissions can still be received succesfully even though they are below the noise floor.

\begin{figure}[ht]
 \centering
 \includegraphics[width=\linewidth,trim=0cm 0cm 0cm 0cm,clip=true]{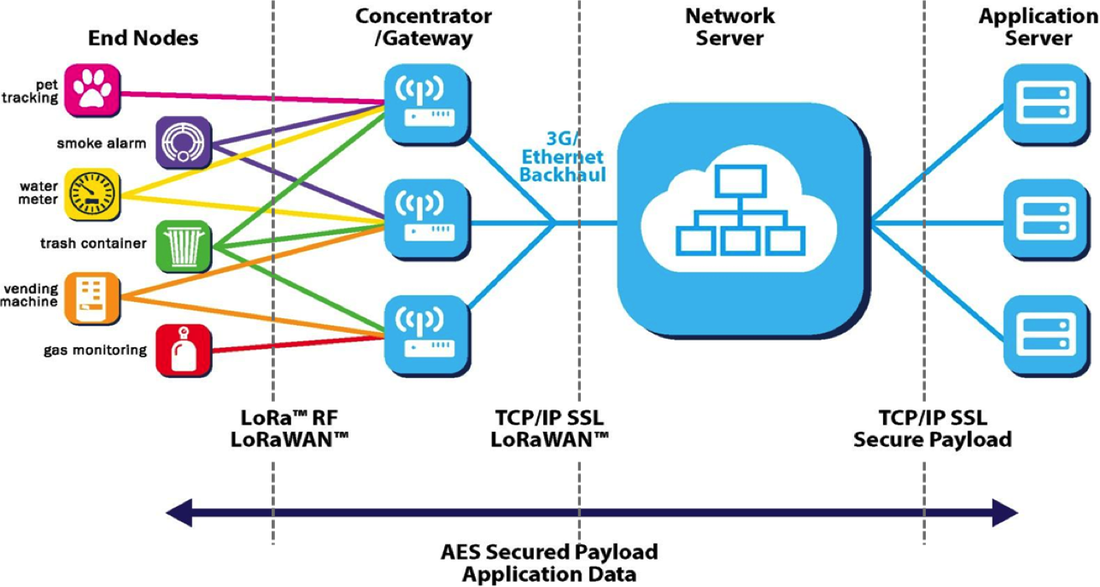}
 \caption{Architecture of LoRaWAN networks (image courtesy of Semtech)}
 \label{fig:lorawan-system}
\end{figure}

LoRaWAN networks employ the robust LoRa modulation in order to achieve long range operation.
They are standardized by the LoRaWAN alliance, which has defined medium access, frame formats, provisioning and management messages, security mechanisms, device management and other aspects.
Figure~\ref{fig:lorawan-system} illustrates that LoRaWAN networks form one hop star topologies around gateways, which act as packet forwarders between end devices and a central network server~(NS).
The network server is responsible for MAC layer processing and acts as a portal between applications running on end devices and application servers.
The LoRaWAN standard defines three classes for end devices~(A, B and C) in order to cater to a number of different scenarios.
This work focuses on class A end devices as this class provides the longest battery life.

Class A end devices have their transceivers in deep sleep for the majority of the time and wake up infrequently to transmit data toward the network server.
Wireless medium access in LoRaWAN follows a ALOHA scheme, which does not employ listen before talk, and is therefor subject to restrictions in most areas of the world.
In Europe for example, the 868MHz band consists of a number of sub-bands where radio duty cycle restrictions range from 0.1\% to 10\% with 1\% being most common.
Note that each of these sub-bands is composed of one or more channels.

\begin{figure}[ht]
 \centering
 \includegraphics[width=0.8\linewidth, trim=0 27.85cm 12.35cm 0, clip=true]{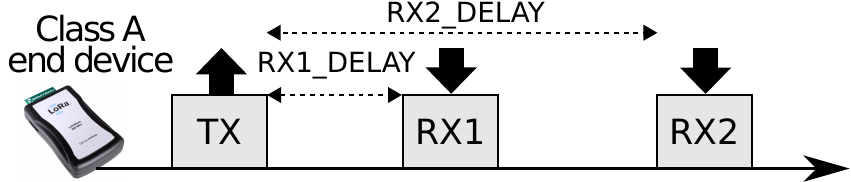}
 \caption{Downlink receive window timing for LoRaWAN class A end devices}
 \label{fig:classA-RW}
\end{figure}

\begin{figure*}[ht!]
 \centering
 \includegraphics[trim=0cm 25.9cm 5.5cm 0cm, clip=true]{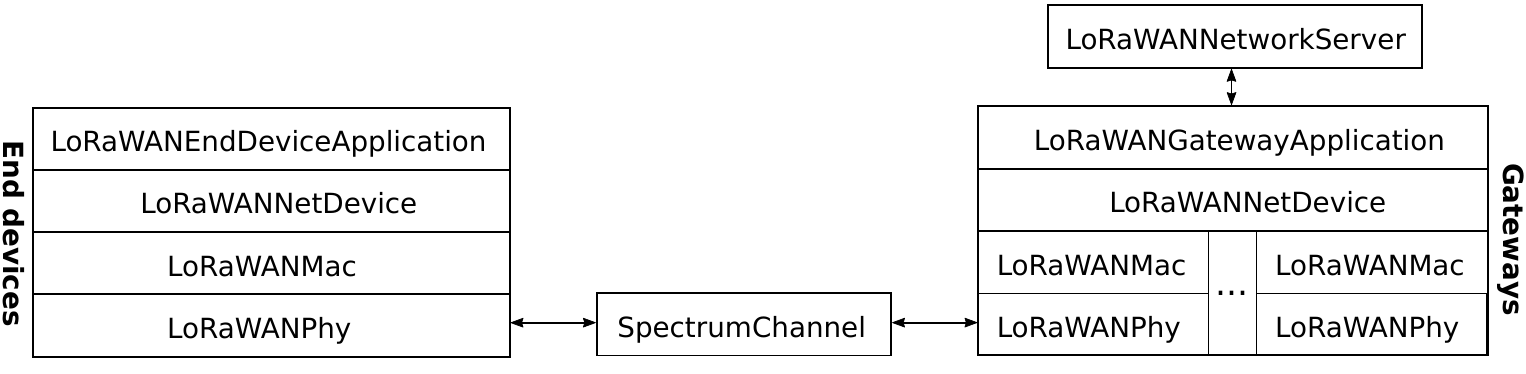}
 \caption{LoRaWAN ns-3 module overview: class A end devices, gateways and the network server}
 \label{fig:lorawan-ns3-module}
\end{figure*}

As class A end devices are unreachable most of the time, the opportunities for sending to the device are scarce.
As per the standard, class A end devices are obliged to open one or two receive windows after each upstream transmission in order to allow the NS to deliver a potential message to the end device.
Figure~\ref{fig:classA-RW} illustrates the timing for opening these windows, which is equal to one and two seconds after the end of the upstream transmission for the first and second receive window respectively.
When an end device receives a downlink transmission in the first window, it is freed from opening the second window. 
Otherwise, it must open the second window.
Note that an end device listens on the same channel and SF as the last upstream transmission in the first receive window~(unless RX1DROffset differs from zero), while it listens on a separate channel and SF12 in the second window.
Additionally, a class A end device must defer all pending upstream transmissions until after the receive window(s).

Finally, both upstream and downstream messages may be sent as either confirmed and unconfirmed messages.
Confirmed messages are sent using a straightforward retransmission scheme at the discretion of the end device, without violating the duty cycle restrictions. 
Downstream (re)transmissions have to wait for an open receive window and their timing is therefor controlled by the upstream traffic timing of an end device.
The NS can however set a frame pending bit in a downstream message in order to signal to an end device that it might want to open a receive window sooner than normal.

To conclude this section ns-3 is briefly introduced. 
Ns-3 is an open source discrete-event network simulator, targeted primarily for research and educational use.
It provides support for Wi-Fi, LTE, 802.15.4 and other networks and also implements an IP networking stack.
A relatively new feature in ns-3 is the SpectrumPhy, which enables modeling of inter-technology interference and which is used in this work to implement the LoRa PHY.

\section{Problem statement and approach}
As mentioned, the anticipated growth of IoT in general and LPWA networks in particular raises the question how these technologies will scale as the size of these networks grows. 
Vendors are keen to highlight the positive aspects of their respective products, but are less inclined to point out certain flaws.
Therefor an objective study of the limits of these LPWA networks is warranted and needed. 

Specifically for LoRaWAN networks, this work attempts to explore the limits of these networks in terms of size. 
This work attempts to answer the following research questions:
\begin{itemize}
 \item What is the impact of end device density on network performance?
 \item What is the impact of the different message types on network performance?
 \item What is the impact of assigning data rates to end devices?
 \item What is gained in terms of network performance by increasing the gateway density?
\end{itemize}

In order to formulate an answer to these questions, a simulation based approach was followed as it allows modeling large scale networks~(i.e. up to 10 000 end devices).
The ns-3 simulator was a natural choice due to its widespread adoption in the network research community.
Also, the new SpectrumPhy feature promises to enable modeling multiple LPWAN technologies in parallel in ns-3 and to study co-existence and other problems in the future.
Finally, this work aspires to provide a useful tool for research into LoRaWAN networks.

\section{LoRaWAN ns-3 module}
\label{sec:lorawan-ns3-module}
Our modeling of LoRaWAN networks in ns-3 comprises a number of different elements.
Firstly, an error model for the LoRa modulation was implemented in ns-3 based on baseband simulations of a LoRa transceiver over an Additive White Gaussian Noise (AWGN) channel.
Secondly, the LoRaWAN PHY and MAC layers were added in ns-3 for gateways and class A end devices.
Thirdly, ns-3 applications were developed to represent LoRaWAN class A end devices and LoRaWAN gateways.
Finally, a simple LoRaWAN network server was added to ns-3.

An overview of the LoRaWAN ns-3 module is presented in figure~\ref{fig:lorawan-ns3-module}.
While end device nodes contain a single MAC/PHY pair, gateways consists of one MAC/PHY pair per supported spreading factor.
For example a gateway that supports multi-SF~(i.e. able to receive all LoRa spreading factors simultaneously) on six channels contains 36 MAC/PHY pairs.
Apart from the components listed in figure, the ns-3 module also contains a number of units tests and examples of varying complexity.
The lorawan ns-3 module is publicly available at \url{https://github.com/imec-idlab/ns-3-dev-git/tree/lorawan}.
Future works that use this ns-3 module, are requested to cite this manuscript.

\subsection{LoRa PHY error model}
\subsubsection{LoRa PHY baseband implementation}
\label{sec:lora-phy-baseband-implementation}
\begin{figure*}[ht]
 \centering
 \includegraphics[width=0.75\textwidth, trim=0cm 25.4cm 3.3cm 0cm, clip]{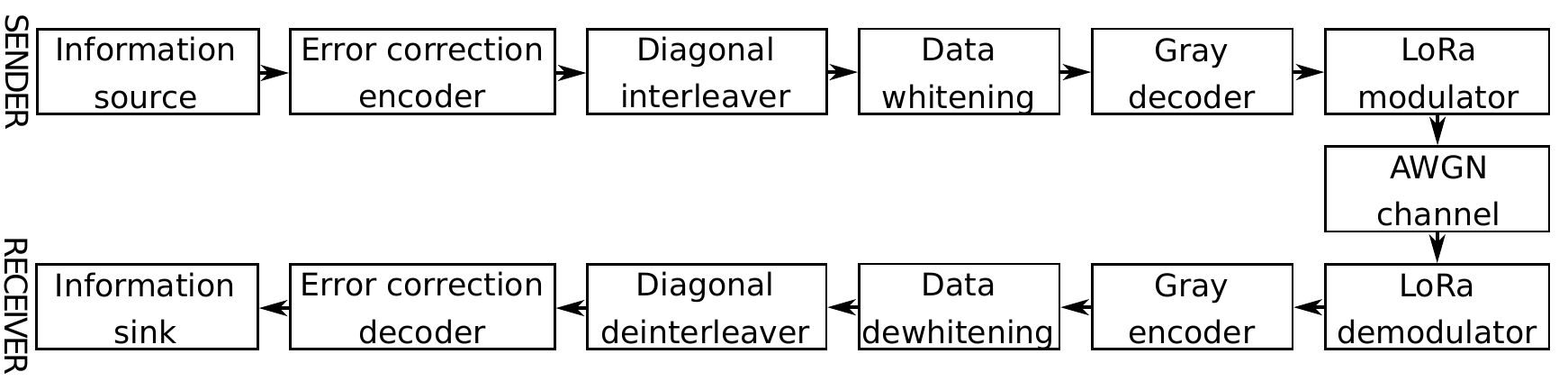}
 \caption{Block diagram of LoRa PHY baseband implementation: sender, AWGN channel and receiver}
 \label{fig:ber-simulations-block-diagram}
\end{figure*}

In order to model the effects of path loss and intra-LoRa interference, an error model for the LoRa PHY has been developed in ns-3.
The basis for this error model is a series of complex baseband Matlab simulations that measure the bit error rate~(BER) for different LoRa PHY configurations over a complex AWGN channel.
A block diagram of the BER simulations is shown in figure~\ref{fig:ber-simulations-block-diagram}.

The information bits generated at the information source are mapped to code bits by the error correction encoder.
This encoder implements the 5/4, 7/4 and 8/4 code rates available in LoRa~\footnote{The 6/4 code rate was not implemented as it is seldom used.}. 
While the 5/4 CR is a simple parity check code, the 7/4 and 8/4 CRs are (7,4) and (8,4) linear error-correcting hamming codes.
These codes can correct one bit error and detect up to two bit errors. 

Next, the diagonal interleaver shuffles the code bits so that at its output, groups of PPM bits consist of bits from the same bit position of PPM consecutive code words.
For example, the first output word groups the bits at position 0 from PPM consecutive code words.
PPM represents the bit length of the output words of the interleaver.
In LoRa the PPM of the interleaver is equal to the LoRa spreading factor.
Consequently, the number of bits mapped per LoRa symbol is equal to the spreading factor.
Due to the interleaver, a lost symbol at the receiver is converted into PPM 1-bit errors over PPM consecutive code words (rather than one PPM-bit error in one code word without the interleaver).

After the interleaver, the output words are whitened in order to boost the entropy of the information source.
Note that in the BER simulations the information bits are drawn from a uniform distribution, therefor the entropy of the information source is already at its maximum.
Before passing the whitened bit stream to the modulator, it is reverse Gray mapped first.
This produces a sequence of integers, which are fed to the LoRa modulator.
At the LoRa modulator, a sequence of N time-shifted complex baseband up-chirp samples is generated via a phase accumulator as given by equation~\ref{eq:modsymbols} where N, the number of samples per baseband symbol, is equal to $\mathit{2^{SF}\frac{f_s}{BW}}$.
The input integer determines the time-shift of the up-chirp.
\begin{equation}\label{eq:modsymbols}
 m(i) = 
  \begin{cases} 
   \exp{(-j \pi)} & \text{if } i = 0 \\
   m(i-1) \exp{(j f(i))}       & \text{if } i = 1,\dots,N-1
  \end{cases}
\end{equation}
Where the instantaneous frequency f(i) is given by equation~\ref{eq:modsymbolsfreq}:
\begin{equation}\label{eq:modsymbolsfreq}
 f(i) = -\pi + \frac{i}{n} 2\pi \text{, for } i = 1,\dots,N-1
\end{equation}

Next, the samples of the LoRa symbol are sent over the AWGN channel for a given signal to noise ratio ($SNR$) as per equation~\ref{eq:awgnchannel}:
\begin{equation}\label{eq:awgnchannel}
\begin{aligned}
c(i) ={} & m(i) + \sqrt{\frac{E_s}{2\mathit{SNR}}}[\mathcal{N}(0;1) + j\mathcal{N}(0;1)]\\
        & \text{for } i = 0,\dots,N-1
\end{aligned}
\end{equation}
where $\mathcal{N}(0;1)$ is the standard normal distribution and $\mathit{SNR}~=~10^{{SNR}_{dB}/10}$.
Note that the energy per symbol is equal to one for the LoRa modulator.

At the receiver, the LoRa demodulator employs correlation-based demodulation where the received symbol is correlated to all known LoRa symbols.
The decision on which symbol was sent, is made by selecting the LoRa symbol with the maximum correlation value.
After demodulation, the receiver chain is the reverse of the sender chain. 
The error rate is measured in the information bits, after error correction.

\subsubsection{LoRa PHY BER simulations}
\label{sec:LoRa-PHY-BER-simulations}
In order to determine the BER of the LoRa physical layer, simulations were ran for the LoRa PHY parameters listed in table~\ref{tab:ber-simulations-lora-phy-params}.
There was no oversampling, so therefor $N = 2^{SF}$ holds.
The simulations were ran for SNR values in steps of 1dB in the ranges as published in the table.
\begin{table}[ht]
\begin{center}
\begin{tabular}{cccc|cccc}
BW & SF & CR & SNR(dB) & BW & SF & CR & SNR(dB)\\
\hline
125kHz & 7 & 1,3 & [-20..0] & 125kHz & 11 & 1 & [-23..-13]\\
125kHz & 8 & 1,3 & [-20..0] & 125kHz & 11 & 3 & [-25..-13]\\
125kHz & 9 & 1,3 & [-20,-8] & 125kHz & 12 & 1,3 & [-26..-17]\\
125kHz & 10 & 1,3 & [-22,-8] &  &  &  & 
\end{tabular}
\end{center}
\caption{LoRa PHY parameters for BER simulations}
\label{tab:ber-simulations-lora-phy-params}
\end{table} 

Afterwards an exponential curve as per equation~\ref{eq:curvefit} was fitted to a subset of the logarithmic values of the measured BER values.
The subset of BER values used for curve fitting was determined as followed. 
Firstly, measured BER values of zero were discarded. 
Secondly, BER values were added to the subset until the BER reached a value where the corresponding packet delivery rate~(PDR) dropped below one in a million for a 13B packet.
The packet length of 13B stems from the minimum LoRa PHY payload length for a LoRaWAN transmission: 1B MAC header, 8B frame header and 4B MIC.
Additionally, for every curve fit a SNR cut-off point was chosen so that the packet delivery rate for a 13B~(=108b) packet was equal to one in a million at the cut-off point.
Table~\ref{tab:ber-simulations-error-model} lists the details of the curve fit plus the SNR cut-off point for every LoRa PHY configuration that was simulated.
\begin{equation}\label{eq:curvefit}
 \log10{(\mathit{BER({SNR}_{dB})})} = \alpha\exp{(\beta{SNR}_{dB})}
\end{equation}
\begin{table}[ht]
\begin{center}
\begin{tabular}{cccccc}
SF & CR & $\alpha$ & $\beta$ & rsquare & SNR cut-off~(dB) \\
\hline
7 & 1 & -30.2580 & 0.2857 & 0.9997  & -12.2833\\
7 & 3 & -105.1966 & 0.3746 & 0.9999 & -12.6962\\
8 & 1 & -77.1002 & 0.2993 & 0.9999 & -14.8485\\
8 & 3 & -289.8133 & 0.3756 & 0.9995 & -15.3588\\
9 & 1 & -244.6424 & 0.3223 & 0.9993 & -17.3749\\
9 & 3 & -1114.3312 & 0.3969 & 0.9994 & -17.9260\\
10 & 1 & -725.9556 & 0.3340 & 0.9996 & -20.0254\\
10 & 3 & -4285.4440 & 0.4116 & 0.9991 & -20.5581\\
11 & 1 & -2109.8064 & 0.3407 & 1.0000 & -22.7568\\
11 & 3 & -20771.6945 & 0.4332 & 0.9996 & -23.1791\\
12 & 1 & -4452.3653 & 0.3317 & 0.9986 & -25.6243\\
12 & 3 & -98658.1166 & 0.4485 & 0.9993 & -25.8602
\end{tabular}
\end{center}
\caption{Exponential curve fit parameters for the LoRa PHY error model in ns-3}
\label{tab:ber-simulations-error-model}
\end{table} 

\subsection{LoRaWAN PHY layer}
After the error model was completed, work started on the implementation of the LoRaWAN Phy layer in ns-3.
By building the LoRaWANPhy class on the SpectrumPhy concept~(introduced in ns-3.26), inter-technology simulations~(e.g. interference testing) are anticipated to be feasible in the future.
The majority of the PHY models available in ns-3 employ a chunk-based signal to interference noise ratio~(SINR) approach for modeling the influence of propagation loss and intra-technology interference during packet reception.
Every time the SINR changes during packet reception~(e.g. an interfering transmission starts or ends), a new chunk is started and the error rate of the previously received chunk is evaluated based on the constant SINR and bit length of this chunk and the BER as provided by the error model.
Note that the LoRaWANPhy only initiates packet reception for transmissions with a SINR value above the SNR cut-off value.
Incoming transmissions which fall below the cut-off value are dropped immediately by the PHY.

\begin{figure}[ht]
 \centering
 \includegraphics[width=0.5\linewidth, trim=1cm 1cm 1cm 1cm, clip]{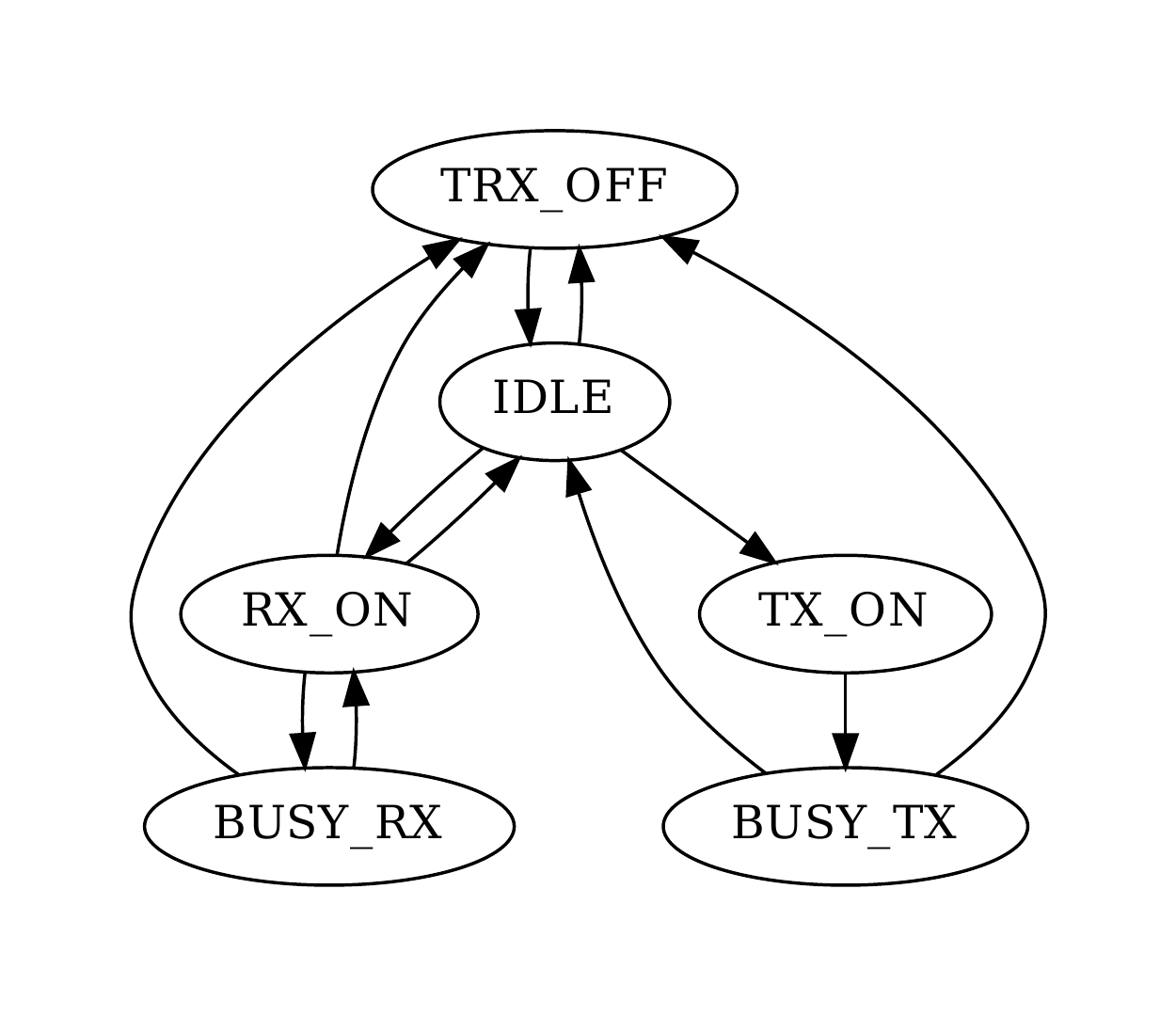}
 \caption{Finite state machine of the LoRaWANPhy class in ns-3}
 \label{fig:lorawan-phy-fsm}
\end{figure}

Apart from the reception modeling, the LoRaWANPhy ns-3 class also implements a finite state machine to structure its execution flow.
The FSM has six states as shown in fig~\ref{fig:lorawan-phy-fsm}.
The transitions between states are mostly triggered from the MAC layer (not shown in the figure).
In the RX\textunderscore ON and TX\textunderscore ON states the PHY is ready to respectively start a packet reception or transmission.
In the BUSY\textunderscore RX state the PHY is busy receiving a transmission~(as per the aforementioned chunk-based reception).
In the BUSY\textunderscore TX state the PHY is sending a transmission.
Ongoing receptions and transmissions may be canceled at any time, which is indicated by the state transitions to TRX\textunderscore OFF.
The PHYs of class A end devices are expected to be in the Idle state most of the time, whereas the PHYs of gateways are expected to be in the RX\textunderscore ON state for the majority of the time.
There are no differences in the PHY ns-3 classes between class A end devices and gateways.
Hence, differences in transceiver design between end devices and gateways are not taken into account in this model.

\subsection{LoRaWAN MAC layer}
The driver of the PHY layer is the LoRaWANMac ns-3 class.
Its functionality includes queuing packets for delivery, opening receive windows and handling retransmissions on end devices and keeping track of a node's radio duty cycle~(RDC).
While there is one LoRaWANMac class for both class A end devices and gateways, the functionality of this class differs as e.g. retransmissions for gateways are handled by the network server~(see section~\ref{sec:lorawan-ns}).
Likewise, the gateway MAC has no concept of receive windows as it always listening for upstream traffic~(when not transmitting).

\begin{figure}[ht]
 \centering
 \includegraphics[width=0.9\linewidth, trim=1.5cm 1.5cm 1.5cm 1.2cm]{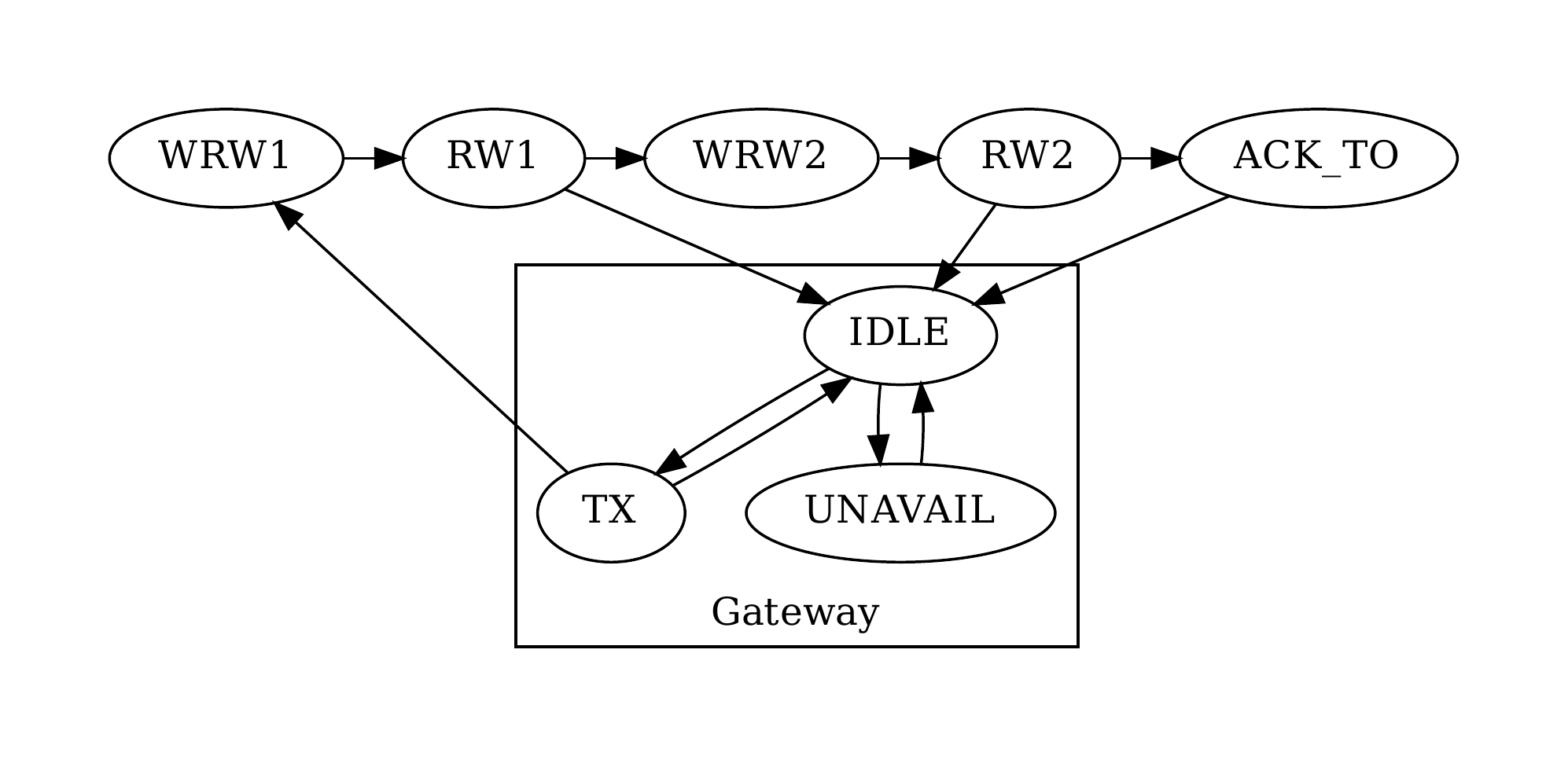}
 \caption{The LoRaWANMac FSM consists of three states for gateways and seven states for class A end devices}
 \label{fig:lorawan-mac-fsm}
\end{figure}

Similarly to LoRaWANPhy, the LoRaWANMac class also implements a FSM as depicted in figure~\ref{fig:lorawan-mac-fsm}.
While an end device MAC object passes through all states in figure~\ref{fig:lorawan-mac-fsm}, gateway MAC objects are limited to three states.
The UNAVAIL state is a case unique to gateways, where the MAC is blocked from sending a packet.
This state is activated when one of the other MACs on the gateway is in the TX state, thereby prohibiting simultaneous transmissions on different MAC objects on the same gateway.

The chain at the top of the figure is related to the mandatory receive windows for class A end devices in LoRaWAN networks.
After the TX state, an end device always transitions to the WRW1 state. 
The end device spends one second~(starting from the end of the transmission) in this `wait for RW1' state, after which it opens RW1.
The end device checks whether a PHY preamble has been received 12.25 LoRa symbols after the beginning of the receive window.
If a preamble has been detected, the device continues receiving the downstream transmission.
If a preamble has not been detected, it closes the receive window and transitions to the WRW2 state.
If the end device successfully receives a downstream transmission in RW1, it transitions to the IDLE state.
Otherwise, it closes the receive window and transitions to the WRW2 state.

In the WRW2 state, the end device is waiting to open the second receive window~(in the RW2 state) after two seconds after the end of the transmission.
The same PHY preamble check from RW1 is performed after opening RW2.
If a downstream transmission is received in RW2, the end device will transition to the IDLE state.
Otherwise, it might transition to the ACK\textunderscore TO~(acknowledgment timeout) state where it spends a random length of time before transiting to the IDLE state.
The ACK\textunderscore TO state is only visited when the end device expected an acknowledgment in one of its receive windows~(i.e. after the transmission of a confirmed upstream message).
In case an acknowledgment is not expected, the device transitions directly to the IDLE state from RW2.

Retransmissions in the case of end devices are handled entirely by the LoRaWANMac class.
As long as the number of remaining transmissions has not expired or a downstream frame with the Ack bit set has not been received, a confirmed data packet will remain in the transmission queue.
The number of transmissions for confirmed messages can be set via \mbox{`DEFAULT\textunderscore NUMBER\textunderscore US\textunderscore TRANSMISSIONS'}~(in lorawan.h) and is set to four by default.
Subsequent (re)transmissions are throttled based on the radio duty cycle limitations of the active sub-band.
To this end there is a per LoRaWAN node singleton object, LoRaWANMacRDC, which keeps track of a node's duty cycle for the different sub-bands.
Finally, the LoRaWANMac class adds a 1B LoRaWANMacHeader~(encoding the message type) and a 4B dummy MIC to the MAC payload before passing the packet on to the PHY.

\subsection{LoRaWAN class A end device ns-3 application}
A new ns-3 application, LoRaWANEndDeviceApplication, was developed to represent class A LoRaWAN end devices in ns-3.
The application exposes attributes for parameters such as the data rate of the end device and the packet length and message type of upstream transmissions.
It also supports configurable random variables for upstream channel selection and packet generation times.
The application is responsible for generating the MAC payload, as such it adds the LoRaWAN frame header to the application payload.
This frame header encodes the end device address, the packet counter and the frame port of the application.
Meta data about the packet transmission - such as the desired channel, data rate and code rate - are passed on to the PHY by means of a LoRaWANPhyParamsTag packet tag.

\subsection{LoRaWAN gateway ns-3 application}
The LoRaWANGatewayApplication is a simple application that is installed on gateway ns-3 nodes.
Apart from passing packets to and accepting packets from the network server, it also supports querying a gateway's RDC status from the NS.
Packets that are to be sent downstream, are tagged with the LoRaWANPhyParamsTag packet tag by the network server.
The LoRaWANNetDevice on the gateway will select the MAC/PHY pair corresponding to the PHY attributes that are listed in the packet tag~(i.e. spreading factor and channel).

\subsection{LoRaWAN Network server}
\label{sec:lorawan-ns}
The LoRaWANNetworkServer class is instantiated only once per LoRaWAN network simulation.
This singleton object accepts upstream packets from gateways and sends downstream traffic to end devices via gateways. 
It exposes the following attributes to configure downstream traffic generation: packet size, confirmed or unconfirmed messages and random variable for packet generation~(an ExponentialRandomVariable by default).
The class keeps track of information such as device address, packet counters, last data rate, last known gateway(s) and last seen time for every end device.
Based on the packet counters, it can detect duplicate data packets from multiple gateways.

The network server generates downstream data and acknowledgments.
To this end, it contains a per end device packet queue for storing downstream traffic.
For every end device, it stores RW1 and RW2 timers that are used for scheduling downstream traffic.
When a timer expires, the network server goes through the list of last known gateway(s) and searches for a gateway that can send the queued downstream packet immediately.
These timers are scheduled every time an upstream transmission is processed by the network server.
Finally, the network server takes care of retransmissions for confirmed downstream data packets.

\section{Scalability analysis of LoRaWAN networks}
\label{sec:scalability-analysis}
The LoRaWAN ns-3 module includes the ``lorawan-tracing-example.cc`` example which was used for all simulations discussed in this section.
It enables automation of simulations from a CLI by setting simulation parameters and outputting ns-3 tracing results to csv files.
The simulations focused on a number of different scenarios, which are detailed in the subsections below.

\begin{figure}[ht]
 \centering
  \includegraphics[trim=0cm 24.7cm 16cm 0cm, clip=true]{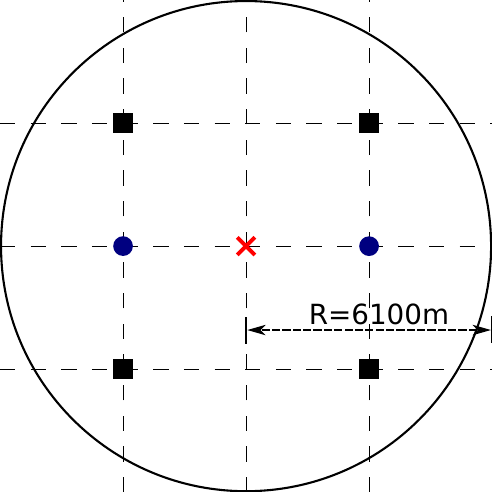}
 \caption{Positions for one~(cross), two~(circles) and four~(rectangles) gateways in ns-3 simulations}
 \label{fig:lorawan-gw-positions}
\end{figure}

All simulations consist of one, two or four gateways and a configurable number of end devices deployed in a disc with a 6 100m radius~\footnote{This radius was chosen as for the presented ns-3 error model, the PDR for 21B packets sent at SF12 lies close to 10\% at this distance for the LogDistancePropagationLoss model.}.
All gateways and end devices are configured to use the same 125kHz LoRaWAN channel~(868.100MHz), with the exception of the high power RW2 channel at 869.525MHz which lies in a sub-band with a 10\% RDC restriction.
This single upstream channel scenario is similar to that of a ``the things gateway'' as sold by The Things Networks.
End devices employ Activation By Personalisation and as such no network join messages are exchanged.
The gateways are deployed at fixed positions, which depend on the number of gateways in the simulation.
In case of one gateway, it is positioned in the origin of the disc.
In case of two gateways, they are positioned one radius apart on a diameter line of the disc.
In case of four gateways, they are positioned on the corners of a square which is centered on the disc origin and which has a diagonal equal to the disc radius. 
The gateway positions are visualized in figure~\ref{fig:lorawan-gw-positions}.
The end devices are uniformly distributed in the disc~(using the UniformDiscPositionAllocator in ns-3) and have a fixed position during the simulation.
For all experiments the default propagation loss model was used in ns-3.
This model, named 'LogDistancePropagationLoss`, has a 3.0 exponent at a 46.6777~dB reference loss at one meter.

Simulation scenarios are run for three different upstream data generation periods: 600, 6 000 and 60 000 seconds.
Each simulation is run for a simulation time equal to hundred times the upstream data generation period.
For every end device, the transmission time of the first upstream packet is picked from a random variable uniformly distributed between zero and the upstream period.
Subsequent upstream packets are periodically generated according to the data generation period. 
Upstream packets have an application payload of 8 bytes, which implies a PHY payload of 21 bytes.

Downstream data generation happens according to an end-device specific exponential random variable~(representing an arrival of events, rather than periodic data transmission).
The mean of this exponential random variable is set to either 60 000s or 600 000s, representing - on average - one downstream packet every ten and hundred upstream packets respectively.

For all simulations, the packet delivery ratio was measured.
An unconfirmed upstream data packet is considered delivered, if it was received successfully by a gateway node.
A confirmed upstream data packet is considered delivered, if one of its transmissions was successfully received by a gateway node and the end device received an acknowledgment from the network server.
Note that the presented PDRs take into account all generated packets, even packets that are queued for transmission are counted towards PDR.
Therefor the PDR reflects overall network throughput~(for the same number of devices and data period).

\subsection{Assigning LoRa spreading factors to end devices}
\label{sec:assigning-lora-spreadings-factors}
The first problem that was studied is how to assign LoRa spreading factors to end devices.
Spreading factors have a major impact on packet delivery rates.
Underestimating the spreading factor~(i.e. assigning a SF that is too low) may lead to reception errors due to low SNR. 
Overestimating the spreading factor~(i.e. assigning a SF that is too high) may lead to inefficient use of air time. 

\begin{figure}[ht]
 \centering
 \includegraphics[width=0.95\linewidth, trim=0.1cm 0.6cm 1cm 1.9cm, clip=true]{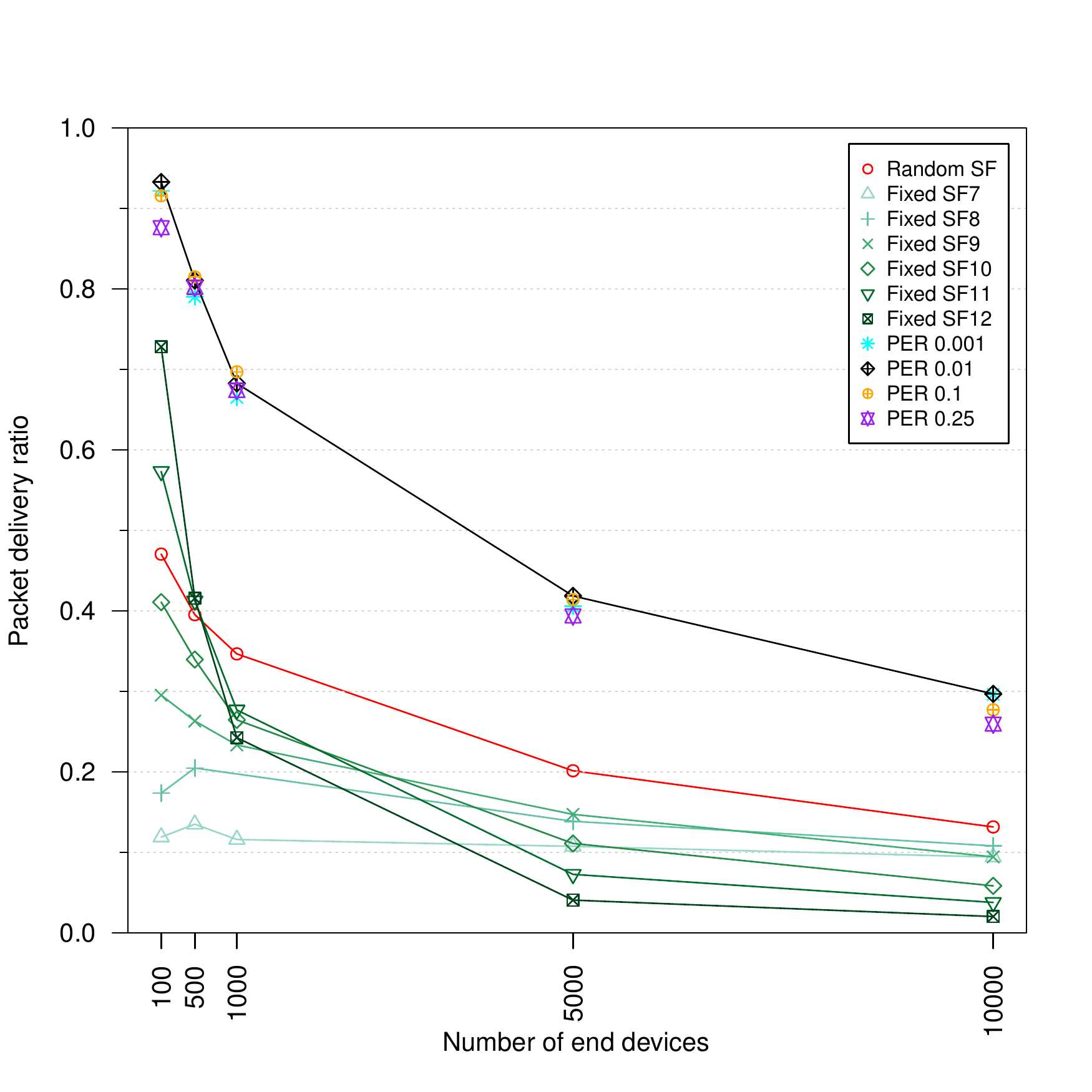}
 \caption{Packet delivery ratios for various spreading factor assignments strategies}
 \label{fig:drcalc-plot}
\end{figure}

Three SF assignment strategies have been considered: 
\begin{enumerate}
 \item Random: assign spreading factors to end devices according to a uniform random distribution.
 \item Fixed: assign the same spreading factor to end devices.
 \item PER: for every end device, find and assign the lowest spreading factor for which the packet error ratio falls below a certain threshold.
\end{enumerate}
For each strategy a number of simulations were performed for a six hundred seconds upstream data period and a varying number of end devices.
For the PER strategy a number of different PER threshold were tested as well: 0.001, 0.01, 0.1 and 0.25.
The packet delivery ratios for the different SF allocation strategies are presented in figure~\ref{fig:drcalc-plot}.
Packet are sent as unconfirmed messages.

\begin{figure}[ht]
 \centering
 \includegraphics[width=\linewidth, trim=0.9cm 1.2cm 1.05cm 2.05cm, clip=true]{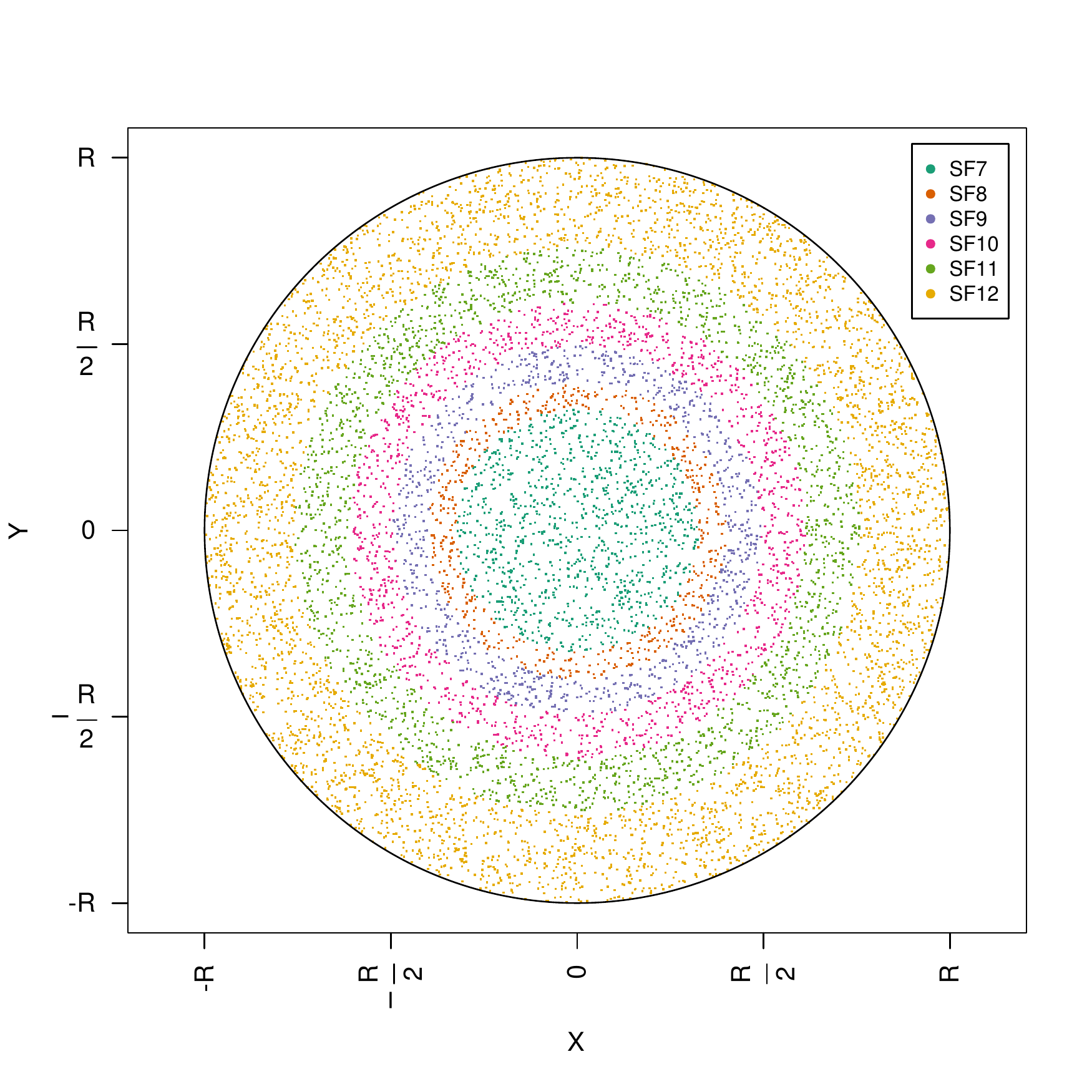}
 \caption{Spreading factor allocation to end devices for PER strategy~(0.01)}
 \label{fig:drcalc-plot-per010-nodes}
\end{figure}

Comparing the results, it is clear that the PER strategy performs the best out of three in terms of PDR.
The PDRs for different PER thresholds are very similar and there is no threshold that yields the highest PDR in all considered network sizes.
Note that while there exist large variations in PDRs between different spreading factors, figure~\ref{fig:drcalc-plot} plots the global PDR across all spreading factors.
A PER threshold of 0.01 is chosen for allocating spreading factors in the remainder of this paper.
With this threshold, there are on average about 43\% SF12, 20\% SF11, 12\% SF10, 8\% SF9, 6\% SF8 and 11\% SF7 in a single gateway LoRaWAN network with radius 6100 meters.
This is represented graphically in figure~\ref{fig:drcalc-plot-per010-nodes}.

\subsection{Unconfirmed vs confirmed upstream data}
\subsubsection{Single gateway LoRaWAN network}
Next, the impact of sending upstream data as confirmed MAC messages is considered on the PDR. 
One would expect the LoRaWAN retransmission scheme to boost the PDR, as unacknowledged messages are retransmitted by the end device. 
An end device attempts four transmissions before dropping the message. 
At all times, end devices respect duty cycle restrictions.

\begin{figure}[ht]
 \centering
 \includegraphics[width=0.95\linewidth, trim=0.15cm 0.6cm 1cm 1.9cm, clip=true]{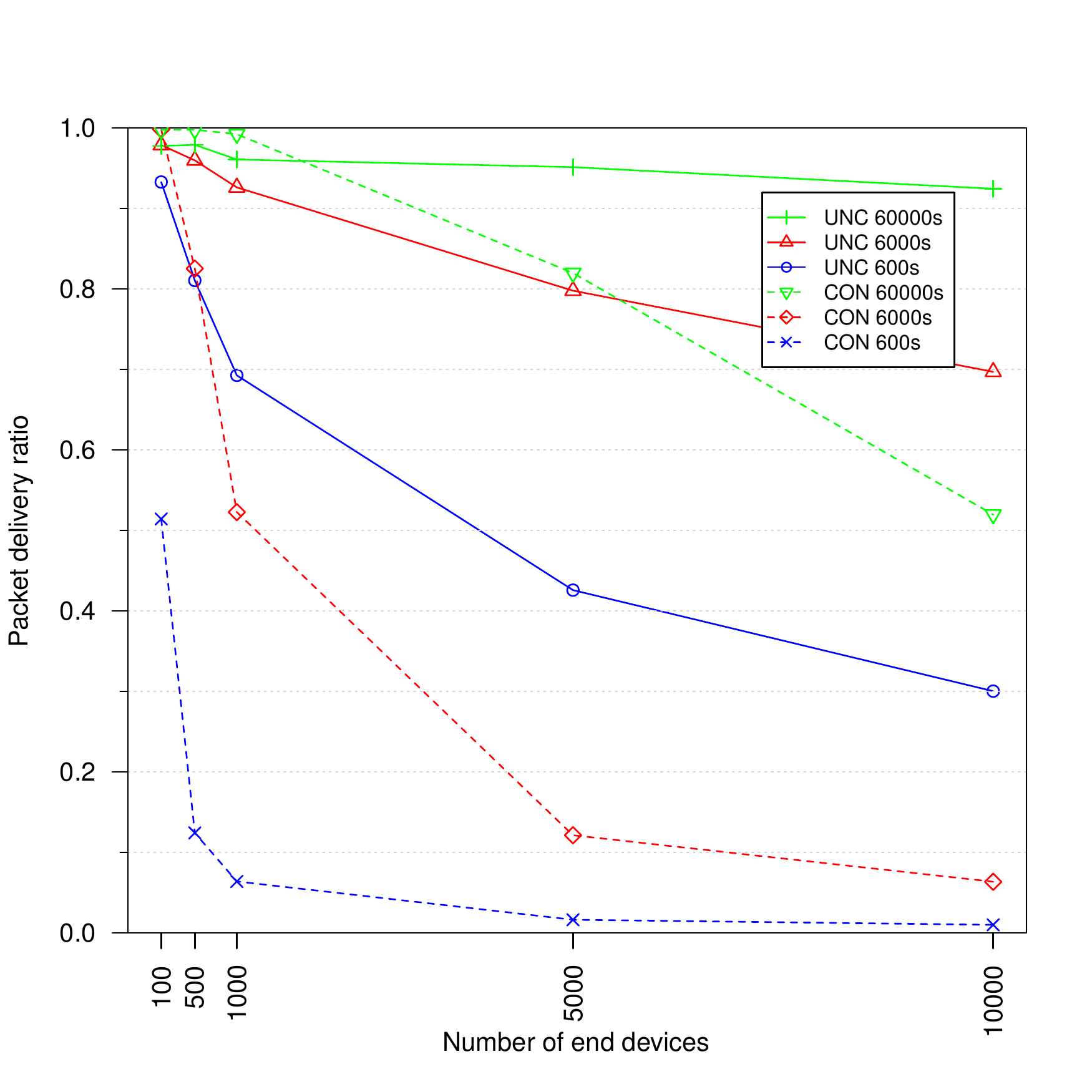}
 \caption{PDR for unconfirmed and confirmed upstream messages in a single gateway LoRaWAN network}
 \label{fig:plot-unconfirmed-confirmed-oneGW}
\end{figure}

The packet delivery ratios for sending upstream data as unconfirmed and confirmed messages are shown in figure~\ref{fig:plot-unconfirmed-confirmed-oneGW} for three different data periods in case of a single gateway LoRaWAN network.
The PDR decreases as data is sent more frequently and as the number of end devices increases.
In case of unconfirmed MAC messages, the primary cause of undelivered packets is due to collisions where the gateway is busy receiving a transmission and therefor any other transmission with the same data rate is dropped during the ongoing reception.
For the 600 seconds data period, the share of drops due to collisions in all undelivered packets is close to 90\%.
Another 9\% of the undelivered packets are destroyed due to interference during reception.
The remainder of the undelivered packets are dropped due to a SINR value that falls below the SNR cut-off point~(cfr. section~\ref{sec:LoRa-PHY-BER-simulations}).

Note that the share of slower data rates in the undelivered packets is higher than that of faster data rates.
This is partially due to the higher share of end devices with slower data rates in the networks and partially due to the higher transmission times at lower data rates. 
The share of undelivered packets sent at SF11 or SF12, lies at 80.9\%, 93.6\% and 95.8\% for the 600, 6~000 and 60~000 seconds data periods respectively.

Somewhat counterintuitively, the PDR of confirmed messages is not always higher than that of unconfirmed messages.
The PDR is only higher in cases where the the traffic load is very low. 
In the simulations this is only the case for 100, 500 and 1000 end devices for a 60~000~s data period and for 100 end devices for a 6~000~s data period. 
In all other cases the PDR of confirmed messages is lower.
Recall that a confirmed message is only considered delivered if the end device receives an acknowledgment for that message.
Table~\ref{tab:rw-stats} shows that the number of missed receive windows~(for sending an acknowledgment) goes up as the traffic load increases.
Receive windows are missed because the gateway is unable to transmit at the start of a receive window due to the duty cycle restrictions that apply in the sub-band of a receive window.
As end devices retransmit more frequently, the average number of packets per confirmed message increases as well.
Finally, when a gateway sends an acknowledgment in either RW1 or RW2 all ongoing receptions at the gateway are aborted; which also impacts PDR. 
\begin{table}[ht]
\begin{center}
\tabcolsep=0.11cm
\begin{tabular}{ccccccc}
\small
GW & DP & \#ED & Ack RW1 & Ack RW2 & Missed RWs & $\mathrm{\frac{\#packets}{message}}$\\
\hline
1 & 60000 & 100 & 8798 & 1354 & 0 \textasteriskcentered\textasteriskcentered\textasteriskcentered & 1.05\\
1 & 60000 & 1000 & 47500 & 53162 & 7968 \textasteriskcentered\textasteriskcentered & 1.19\\
1 & 60000 & 10000 & 91950 & 438343 & 1604427 \textasteriskcentered & 3.08\\
1 & 6000 & 100 & 4741 & 5316 & 1122 & 1.21\\
1 & 6000 & 1000 & 9542 & 43767 & 155513 & 3.07\\
1 & 6000 & 10000 & 17078 & 52033 & 1465697 & 3.90\\
1 & 600 & 100 & 943 & 4315 & 15052 \textdagger\textdagger\textdagger & 3.08\\
1 & 600 & 1000 & 1623 & 5199 & 143153 \textdagger\textdagger & 3.90\\
1 & 600 & 10000 & 6880 & 5273 & 262385 \textdagger & 3.98\\
\hline
2 & 60000 & 100 & 9896 & 174 & 0 \textasteriskcentered\textasteriskcentered\textasteriskcentered & 1.02\\
2 & 60000 & 1000 & 75062 & 24590 & 1026 \textasteriskcentered\textasteriskcentered & 1.07\\
2 & 60000 & 10000 & 248682 & 631806 & 877387 \textasteriskcentered & 2.09\\
2 & 6000 & 100 & 7926 & 2170 & 0 & 1.03\\
2 & 6000 & 1000 & 25400 & 63095 & 84263 & 2.04\\
2 & 6000 & 10000 & 42798 & 103371 & 2229616 & 3.79\\
2 & 600 & 100 & 2513 & 6477 & 8502 \textdagger\textdagger\textdagger & 2.02\\
2 & 600 & 1000 & 4838 & 10327 & 216539 \textdagger\textdagger & 3.77\\
2 & 600 & 10000 & 13117 & 10577 & 645143 \textdagger & 3.97\\
\hline
4 & 60000 & 100 & 10012 & 0 & 0 \textasteriskcentered\textasteriskcentered\textasteriskcentered & 1.00\\
4 & 60000 & 1000 & 95350 & 4866 & 201 \textasteriskcentered\textasteriskcentered & 1.01\\
4 & 60000 & 10000 & 646058 & 355183 & 128053 \textasteriskcentered & 1.17\\
4 & 6000 & 100 & 9380 & 656 & 2 & 1.00\\
4 & 6000 & 1000 & 66712 & 33302 & 12972 & 1.16\\
4 & 6000 & 10000 & 135780 & 201664 & 2433131 & 3.47\\
4 & 600 & 100 & 6568 & 3470 & 1360 \textdagger\textdagger\textdagger & 1.15\\
4 & 600 & 1000 & 14906 & 20085 & 242954 \textdagger\textdagger & 3.45\\
4 & 600 & 10000 & 26866 & 21163 & 1306882 \textdagger & 3.94\\
\end{tabular}
\end{center}
\caption{Transmission of acknowledgments for one, two and four gateway LoRaWAN network simulations}
\label{tab:rw-stats}
\end{table} 

\subsubsection{Multi gateway LoRaWAN networks}
\begin{table}[h!]
\begin{center}
\begin{tabular}{c|cccccc}
GW & SF7 & SF8 & SF9 & SF10 & SF11 & SF12 \\
\hline
1 & 11\% & 6\% & 8\% & 12\% & 20\% & 43\%\\
2 & 21\% & 10\% & 17\% & 18\% & 16\% & 18\%\\
4 & 40\% & 16\% & 23\% & 17\% & 4\% & 0\\
\end{tabular}
\end{center}
\caption{End devices at a specific data rate for LoRaWAN networks with one, two and four gateways}
\label{tab:data-rates-vs-nr-gateways}
\end{table} 

In this section the effect of the number of gateways in a LoRaWAN network on the PDR is studied. 
Note that for applying the PER 0.01 SF allocation strategy, the PER to the closest gateway is calculated for every end device.
Increasing the gateway density, as per figure~\ref{fig:lorawan-gw-positions}, is anticipated to have more than one effect.
Firstly, it should enable higher data rates for end devices due to an increase in link budget~(as on average gateways will appear closer).
Secondly, as downstream transmissions in RW1 are sent with the same data rate as the upstream transmissions, RW1 acknowledgments should also profit from the higher data rates of end devices.
Finally, as duty cycle restrictions apply per gateway, the LoRaWAN network should be able to acknowledge more messages as the gateway density goes up.
\begin{figure}[ht]
 \centering
 \includegraphics[width=0.95\linewidth, trim=0.15cm 0.6cm 1cm 1.9cm, clip=true]{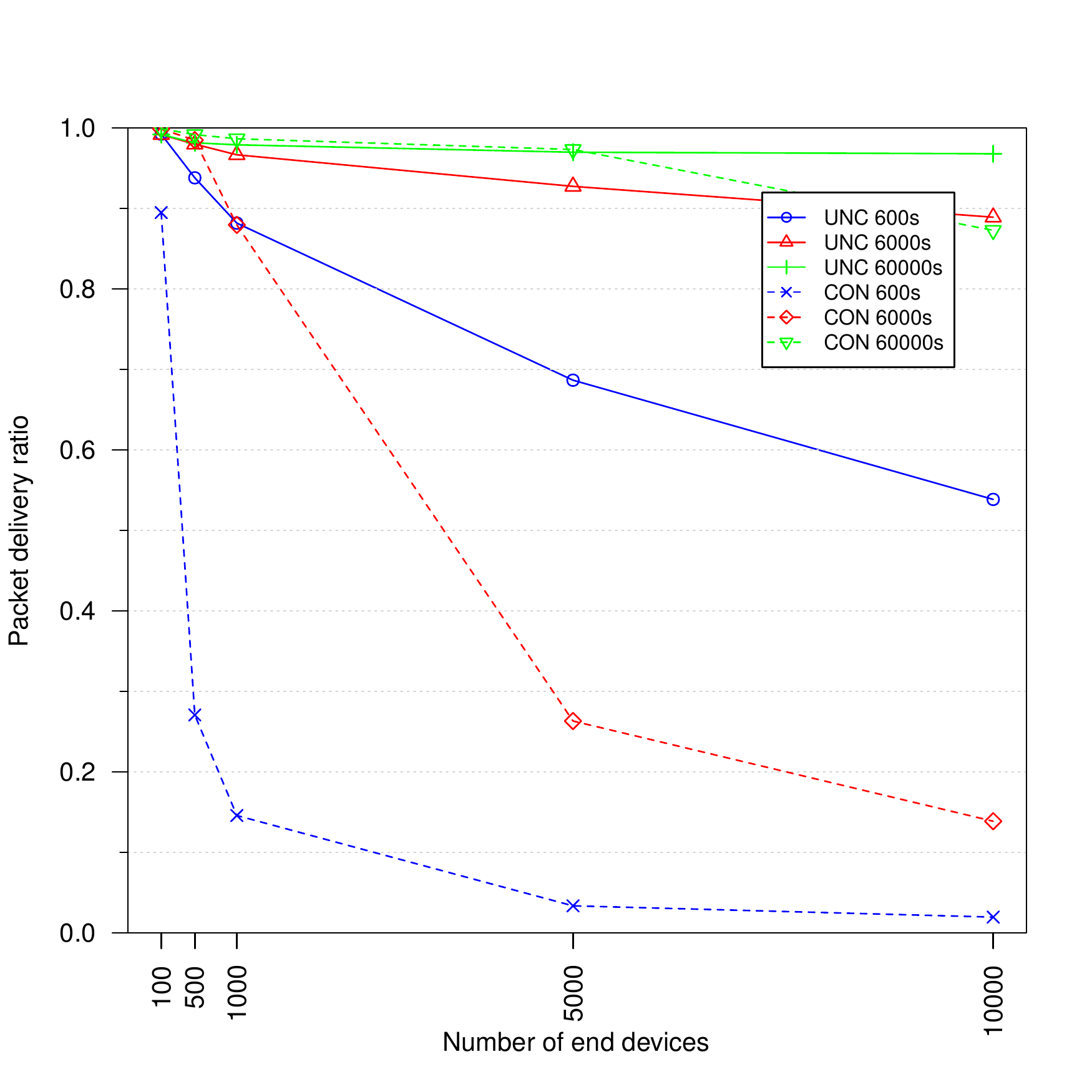}
 \caption{PDR for unconfirmed and confirmed upstream messages in a two gateway LoRaWAN network}
 \label{fig:plot-unconfirmed-confirmed-twoGW}
\end{figure}

Table~\ref{tab:data-rates-vs-nr-gateways} lists the fraction of end devices at specific data rates in a 10 000 end devices LoRaWAN network with one, two and four gateways~(following the PER SF assignment strategy, see section~\ref{sec:assigning-lora-spreadings-factors}).
The table clearly illustrates that higher gateway densities lead to faster overall data rates.

\begin{figure}[hb]
 \centering
 \includegraphics[width=0.95\linewidth, trim=0.15cm 0.6cm 1cm 1.9cm, clip=true]{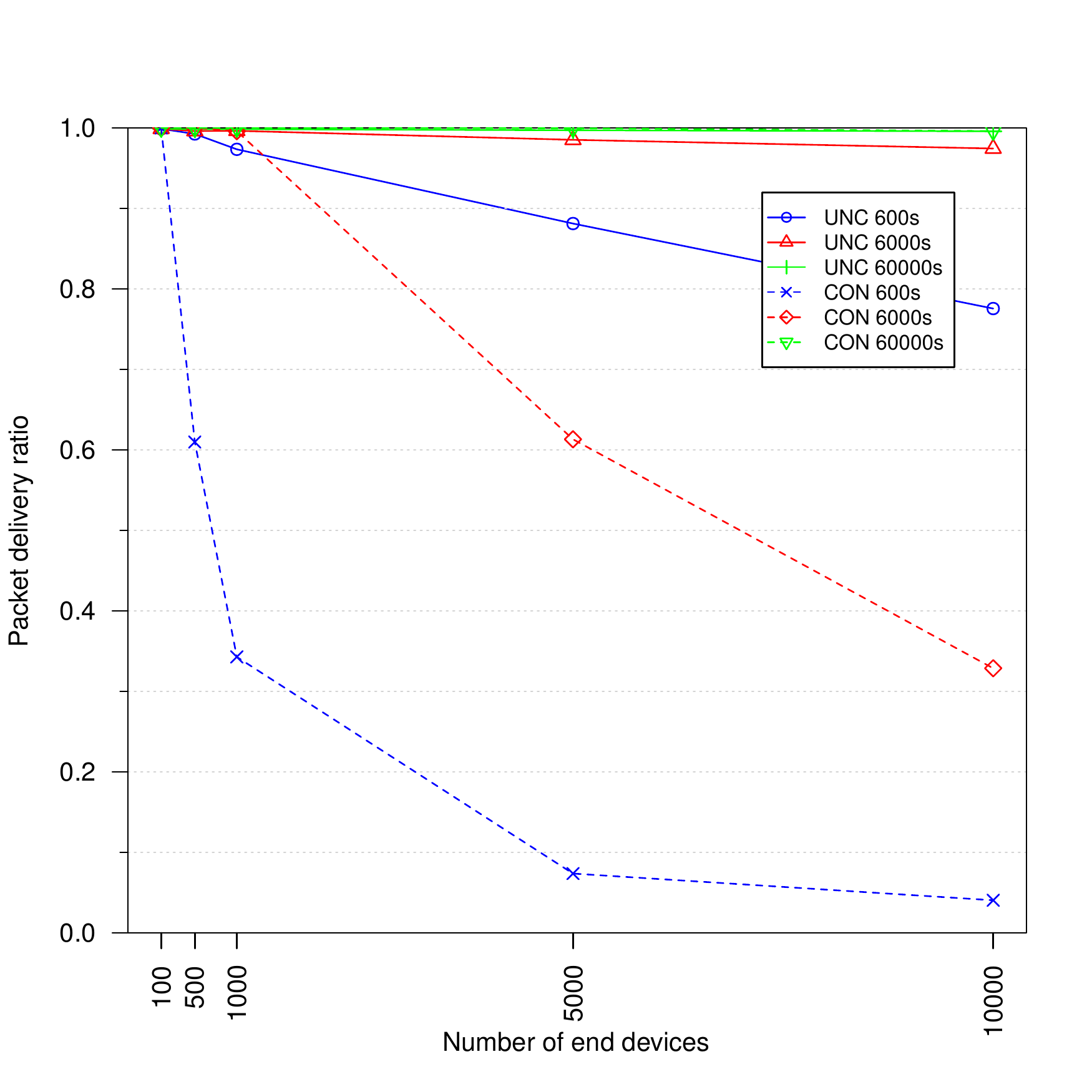}
 \caption{PDR for unconfirmed and confirmed upstream messages in a four gateway LoRaWAN network}
 \label{fig:plot-unconfirmed-confirmed-fourGW}
\end{figure}

Figures~\ref{fig:plot-unconfirmed-confirmed-twoGW} and~\ref{fig:plot-unconfirmed-confirmed-fourGW} show the PDR for a LoRaWAN network with two and four gateways respectively.
Notice how for unconfirmed messages, the PDR increases greatly as the number of gateways increases.
For confirmed messages, the increase in PDR is noticeable but is is not as sharp as for unconfirmed messages.
Studying table~\ref{tab:rw-stats}, it is clear that the number of sent acknowledgments increases as the number of gateways increases.
The seemingly contradicting relation between number of missed RWs and the number of gateways is explained as follows. 
In saturated LoRaWAN networks~(i.e. scenarios with low PDRs for unconfirmed messages), the number of sent messages that are successfully received increases with the gateway density.
In case of confirmed messages, the higher number of received upstream messages means that the network server is able to identify a larger number of receive windows of end devices~(as RWs are always opened after a transmission of an end device).
When gateways are unable to sent in these receive windows~(due to duty cycle restrictions), the number of missed RWs increases.
This is illustrated for the simulation scenarios marked with the \textdagger~symbol in table~\ref{tab:rw-stats}.
In less saturated scenarios~(marked with the \textasteriskcentered~symbol), the number of missed RWs goes down as increasing the gateway density does not lead to identifying more receive windows.
Instead, the number of missed RWs decreases and more acknowledgments are sent~(as seen in columns RW1 and RW2), which benefits the PDR.

\subsection{Downstream data traffic}
\label{sec:ds-traffic}
In the final part of this evaluation, the impact of sending downstream data is studied.
While most LoRaWAN deployments are expected to exhibit high asymmetry between the volume of upstream and downstream data, occasional downstream data messages are expected to be sent.
Potential reasons for downstream data include notifying the end device of an event, end device and network management and updating application parameters~(e.g. sensor sampling interval).
Due to the sparseness and stochastic nature~(e.g. events) of downstream data, generation of downstream data messages in ns-3 is modeled via a per-end device Poisson process with a configurable average rate~$\mathrm{\lambda}$ and mean inter arrival time~$\mathrm{\mu=\frac{1}{\lambda}}$.

In terms of simulations the upstream scenario with a data period of six thousand seconds is chosen as a starting point.
Two average downstream rates of one DS packet every 60 000s and 600 000s are considered, which corresponds to one DS packet every ten and hundred US packets respectively.
Both confirmed and unconfirmed downstream messages and confirmed and unconfirmed upstream messages are taken into account.
Simulations were ran for one, two and four gateway networks with 100, 500, 1 000, 5 000 and 10 000 end devices.
Downstream packets have a 21B size, which holds eight bytes of application payload.

\begin{table}[ht]
\centering
\tabcolsep=0.19cm
\begin{tabular}{clllll|lllllc}
\multicolumn{12}{c}{\textbf{Downstream Packet Delivery Ratios with 1 GW}}\\
                    US               & \multicolumn{5}{c}{DS UNC} & \multicolumn{5}{c}{DS CON} & \multicolumn{1}{c}{$\mathrm{\mu}$} \\
\multirow{2}{*}{\rotatebox{90}{UNC}} & 98 & 97 & 94 & 67 & 40 & 99 & 97 & 93 & 59 & 33 & 10\\
                                     & 99 & 96 & 92 & 80 & 69 & 100 & 96 & 93 & 80 & 69 & 100\\\cline{2-11}
\multirow{2}{*}{\rotatebox{90}{CON}} & 100 & 98 & 92 & 50 & 31 & 100 & 97 & 90 & 48 & 30 & 10\\
                                     & 99 & 98 & 93 & 63 & 45 & 100 & 97 & 92 & 61 & 44 & 100\\
                        
\multicolumn{12}{c}{\textbf{Downstream Packet Delivery Ratios with 2 GWs}}\\
                                     & \multicolumn{5}{c}{DS UNC} & \multicolumn{5}{c}{DS CON} & \multicolumn{1}{c}{$\mathrm{\mu}$} \\
\multirow{2}{*}{\rotatebox{90}{UNC}} & 100 & 98 & 97 & 91 & 75 & 99 & 97 & 96 & 89 & 68 & 10\\
                                     & 100 & 98 & 97 & 93 & 88 & 99 & 97 & 97 & 92 & 87 & 100\\\cline{2-11}
\multirow{2}{*}{\rotatebox{90}{CON}} & 100 & 99 & 98 & 79 & 59 & 99 & 99 & 98 & 78 & 58 & 10\\
                                     & 99 & 99 & 98 & 85 & 73 & 99 & 100 & 98 & 84 & 72 & 100\\
                        
\multicolumn{12}{c}{\textbf{Downstream Packet Delivery Ratios with 4 GWs}}\\
                                     & \multicolumn{5}{c}{DS UNC} & \multicolumn{5}{c}{DS CON} & \multicolumn{1}{c}{$\mathrm{\mu}$} \\
\multirow{2}{*}{\rotatebox{90}{UNC}} & 100 & 100 & 100 & 98 & 96 & 100 & 99 & 99 & 98 & 96 & 10\\
                                     & 100 & 100 & 100 & 98 & 97 & 100 & 100 & 100 & 97 & 96 & 100\\\cline{2-11}
\multirow{2}{*}{\rotatebox{90}{CON}} & 100 & 100 & 100 & 97 & 91 & 100 & 99 & 99 & 97 & 88 & 10\\
                                     & 100 & 100 & 100 & 97 & 94 & 100 & 100 & 100 & 97 & 92 & 100\\
\end{tabular}
\vspace{1mm}
\caption{Packet delivery ratios of downstream data messages}
\label{tab:downstream-simulations-PDR-DS}
\end{table}

Tables~\ref{tab:downstream-simulations-PDR-DS} and~\ref{tab:downstream-simulations-PDR-US} present the PDRs of downstream and upstream data messages respectively for the different parameters that were tested.
The five columns per quadrant in the tables represent results for 100, 500, 1 000, 5 000 and 10 000 end devices from left to right.

\begin{table}[ht]
\centering
\tabcolsep=0.19cm
\begin{tabular}{clllll|lllllc}
\multicolumn{12}{c}{\textbf{Upstream Packet Delivery Ratios with 1 GW}}\\
                US                   & \multicolumn{5}{c}{DS UNC} & \multicolumn{5}{c}{DS CON} & \multicolumn{1}{c}{$\mathrm{\mu}$} \\
\multirow{2}{*}{\rotatebox{90}{UNC}} & 98 & 95 & 90 & 70 & 60 & 98 & 95 & 89 & 70 & 60 & 10\\
                                     & 98 & 96 & 92 & 79 & 68 & 98 & 96 & 92 & 79 & 68 & 100\\\cline{2-11}
\multirow{2}{*}{\rotatebox{90}{CON}} & 100 & 81 & 49 & 10 & 5 & 100 & 81 & 49 & 10 & 5 & 10\\
                                     & 100 & 83 & 52 & 12 & 6 & 100 & 83 & 52 & 12 & 6 & 100\\
                        
\multicolumn{12}{c}{\textbf{Upstream Packet Delivery Ratios with 2 GWs}}\\
                                     & \multicolumn{5}{c}{DS UNC} & \multicolumn{5}{c}{DS CON} & \multicolumn{1}{c}{$\mathrm{\mu}$} \\
\multirow{2}{*}{\rotatebox{90}{UNC}} & 99 & 98 & 96 & 87 & 80 & 99 & 97 & 96 & 87 & 80 & 10\\
                                     & 99 & 98 & 97 & 92 & 88 & 99 & 98 & 97 & 92 & 88 & 100\\\cline{2-11}
\multirow{2}{*}{\rotatebox{90}{CON}} & 100 & 98 & 87 & 23 & 11 & 100 & 98 & 87 & 23 & 11 & 10\\
                                     & 100 & 98 & 88 & 26 & 14 & 100 & 98 & 88 & 26 & 13 & 100\\
                        
\multicolumn{12}{c}{\textbf{Upstream Packet Delivery Ratios with 4 GWs}}\\
                                     & \multicolumn{5}{c}{DS UNC} & \multicolumn{5}{c}{DS CON} & \multicolumn{1}{c}{$\mathrm{\mu}$} \\
\multirow{2}{*}{\rotatebox{90}{UNC}} & 100 & 100 & 99 & 97 & 94 & 100 & 100 & 99 & 97 & 94 & 10\\
                                     & 100 & 100 & 100 & 98 & 97 & 100 & 100 & 100 & 98 & 97 & 100\\\cline{2-11}
\multirow{2}{*}{\rotatebox{90}{CON}} & 100 & 100 & 100 & 58 & 30 & 100 & 100 & 100 & 58 & 29 & 10\\
                                     & 100 & 100 & 100 & 61 & 33 & 100 & 100 & 100 & 61 & 33 & 100\\
\end{tabular}
\vspace{1mm}
\caption{Packet delivery ratios of upstream messages in the presence of downstream data}
\label{tab:downstream-simulations-PDR-US}
\end{table}

Studying table~\ref{tab:downstream-simulations-PDR-DS} the effect of saturating the available airtime at the gateway is clearly visible for simulations with one gateway and a large number of nodes~(i.e. $\geq$5000). 
As the number of gateways increases, the downstream traffic load is spread over more gateway which leads to less saturation per gateway and therefor to an increase in downstream PDR.
The numbers also show that sending US data as confirmed messages negatively impacts the PDR of downstream data messages.
This is because confirmed US messages require a downstream message for an acknowledgment, which increases the traffic load and therefor saturation on the gateway(s).
Finally, table~\ref{tab:downstream-simulations-PDR-DS} also shows that for saturated scenarios the PDR for confirmed downstream messages is slightly lower than for unconfirmed downstream messages.
While the cause of this is not obvious, it is probable that the 70-80\% PDR of US messages in saturated scenarios~(see UNC 6000s figure~\ref{fig:plot-unconfirmed-confirmed-oneGW}) leads to losses of upstream acknowledgments which decreases downstream data PDR in the case of confirmed DS messages.

Comparing table~\ref{tab:downstream-simulations-PDR-US} to the 6000s US PDRs in figures~\ref{fig:plot-unconfirmed-confirmed-oneGW},~\ref{fig:plot-unconfirmed-confirmed-twoGW} and~\ref{fig:plot-unconfirmed-confirmed-fourGW}, the presence of DS data traffic leads to a negligible decrease in US PDR for low DS traffic rates~($\mathrm{\mu=100}$) and a small decrease in US PDR for the high DS traffic rate~($\mathrm{\mu=10}$) for scenarios with 5 000 and 10 000 end devices.
The decrease is more profound for unconfirmed US messages than confirmed US messages, which indicates an increase in US packet loss.
This increase is US packet loss is due to the gateway being unable to receive US transmissions during a DS transmission.
As more gateways are deployed, the DS data transmissions occupy less time per gateway~(due to overall higher data rates) which means that the gateways can spent more time on listening for US messages, thereby reducing the effect of DS data traffic on US packet loss.
Finally, note that there is no difference in terms of US PDR between confirmed and unconfirmed downstream data messages.

\section{Related work}
\label{sec:related-work}
A number of works have been published in literature that study the scalability of LoRa(WAN) LPWA networks.

In one of the first works on this topic, Mikhaylov et al.~\cite{7499263} present an analysis of the capacity and scalability of LoRa LPWANs.
The authors perform an analytical analysis of the maximum throughput for a single LoRaWAN end device, taking into account such factors as RDC and the influence of receive windows.
The authors note that receive windows drastically increase the time between subsequent transmissions and that RDC restrictions reduce the maximum throughput further.
The authors applied the same methodology to determine the capacity of LoRaWAN based on ALOHA access.
While it is true that the LoRaWAN MAC access is an ALOHA scheme, empirical data has shown that the assumptions made in pure ALOHA access do not adequately model a LoRaWAN network~(see figure 4 in~\cite{Bor2016a}). 
Specifically, it fails to model the interference between concurrent transmissions as pure ALOHA assumes concurrent transmissions are always lost regardless of their received power levels, timings and the presence of forward error correction.
A second, but similarly lacking, pure ALOHA capacity analysis of LoRaWAN is discussed in~\cite{s16091466}.
In~\cite{adelantado2017understanding} Adelantado et al. also calculate LoRaWAN capacity as the superposition of independent ALOHA-based networks~(one for each channel and for each SF).
In conclusion, analyses based on pure ALOHA, fail to adequately model interference in LoRaWAN networks and therefor underestimate the capacity of LoRaWAN LPWANs.

In~\cite{georgiou2017low}, Georgiou and Raza provide a stochastic geometry framework for modeling the performance of a single channel LoRa network.
Two independent link-outage conditions are studied, one which is related to SNR~(i.e. range) and another one which is related to co-spreading factor interference.
The authors argue that LoRa networks will inevitably become interference-limited, as end device coverage probability decays exponentially with increasing number of end devices.
The authors report that this is mostly caused by co-spreading factor interference and that the low duty cycle and chirp orthogonality found in LoRa do little to mitigate this.
Finally, the authors note that the lack of a packet-level software simulation is hindering the study into the performance of LoRa.
It would be interesting to combine the authors' modeling of co-spreading factor interference with our ns-3 error model, as in the SINR approach all interference is treated as noise.

The work of Bor et al.~\cite{Bor2016a} studies the limit on the number of transmitters supported by a LoRa system based on an empirical model.
The authors performed practical experiments that quantify communication range and capture effect of LoRa transmissions.
These findings were used to build a purpose-built simulator, LoRaSim, with the goal of studying the scalability of LoRa networks.
The authors conclude that LoRa networks can scale quite well if they use dynamic transmissions parameter selection and/or multiple sinks.
Our study confirms that multiple sinks drastically improve scalability, even though we use a very different approach for modeling interference.
Furthermore, our study goes deeper into modeling LoRaWAN as the LoRaWAN MAC layer is modeled and the impact of confirmed messages and downstream traffic is studied.

The recent work presented by Pop et al. in~\cite{PopBidirectional2017} studies the impact of bidirectional traffic in LoRaWAN by extending the LoRaSim simulator to include bidirectional LoRaWAN communication. 
The resulting simulator is named LoRaWANSim.
Both our ns-3 module and LoRaWANSim allow to study the scalability of LoRaWAN networks.
Both works find that duty cycle limitations at the gateway limit the number of downlink messages~(Ack or data) a gateway can send.
This problem grows worse as the end device density increases, but can be partially mitigated by increasing gateway density~(see section~\ref{sec:ds-traffic}).
The authors of~\cite{PopBidirectional2017} correctly identify that the absence of an acknowledgement, does not necessarily mean that the link quality has decreased and that a node should decrease its data rate for subsequent retransmissions.
Actually, decreasing the data rate might exacerbate this problem as detailed in~\cite{PopBidirectional2017}.
Notable differences between the two simulators include that the LoRaWANSim manuscript is limited to single gateway network, while the ns-3 module provides support for multi-gateway LoRaWAN networks.
Secondly, the collision models are quite different. 
The ns-3 module builds on the error model derived from the complex baseband BER simulations, while LoRaWANSim reuses the empirical model from LoRaSim.
Both collision models support the capture effect as well as modeling interference.
Under capture effect, we understand the ability to receive an interfered transmission in the presence of one or more interferers as long as the SNR of the interfered transmission is sufficiently high for the transmission to be received error-free.
The LoRaWANSim collision model incorrectly assumes perfect orthogonality between spreading factors, while the ns-3 module counts every transmission on the same channel with a different spreading factor as interference.
Furthermore, the LoRaWANSim manuscript does not mention the 10\% RDC restriction that applies in the sub-band of the RW2 channel in the EU. 
This underestimates the downlink capacity in RW2.
Thirdly, the SpectrumPhy model for the LoRa PHY in ns-3 enables modeling inter-technology interference, which could facilitate studies on the interference between 802.11ah on LoRaWAN.
Finally, the LoRaWANSim simulator does not appear to be open source although the manuscript is still under revision at this time.


\section{Discussion}
\label{sec:discussion}

In this section a number of findings from our scalability analysis in section~\ref{sec:scalability-analysis} are discussed.
The results show that confirmed messages severely impact the packet delivery ratios of upstream messages.
While increasing the number of gateways helps to alleviate this problem somewhat, the results of the six hundred seconds data period show that the PDR remains low even in a four gateway network. 
The impact of downstream data messages on upstream messages was found to be negligible due to the sparseness of the tested downstream data traffic load.
Additionally, little difference was found between sending downstream data as unconfirmed messages vs confirmed messages in terms of the DS PDR.
Only for the single gateway and $\mathit{\mu~=~10}$ scenario a significant difference was found.

As every study has its limitations, a number of points that could be improved as part of future work are discussed here.
As discussed in the related work study, the approach of modeling all interference as noise has its drawbacks.
Specifically, literature has shown that while interference between different spreading factors can be accurately modeled as noise, co-spreading factor interference may be modeled more accurately via a stochastic approach.
Future studies may opt to fine-tune the path loss model in ns-3 in order to more closely match the radio environment under study.
An interesting point for future work is to study the impact of the downstream data rate in RW2.
By default, this is set to the lowest data rate in the LoRaWAN standard.
However when downstream data messages are not delivered due to RDC limitations~(rather than low link quality), a faster RW2 data rate might increase the capacity of the LoRaWAN.
Another interesting research topic would be to introduce structure to the LoRaWAN medium access.
While this will come at a cost in terms of traffic overhead and power consumption, it might lead to higher network capacity by reducing interference. 
Additional MAC features such as adaptive data rate~(ADR) and network management~(e.g. joining) could be added to the ns-3 module.
This would allow a more in-depth study of the LoRaWAN standard.
Finally, it would be interesting to study how LoRaWAN networks are affected by the presence of other sub-GHz (LP)WAN technologies.

\section{Conclusion}
\label{sec:conclusion}

In this work a comprehensive model of LoRaWAN LPWANs in the ns-3 network simulator is presented.
This model includes an error model used for determining range as well as interference between multiple simultaneous transmissions.
All spreading factors and coding rates found in LoRaWAN are supported by the PHY layer model of LoRa in the ns-3 module.
The ns-3 module models the MAC layer for class A end devices and supports both upstream and downstream (un)confirmed messages via a simple network server.
Furthermore, LoRaWAN networks with multiples gateways are supported.

The ns-3 module forms the basis for a scalability analysis of single channel multi gateway LoRaWAN LPWANs.
The results of this analysis show that allocating network parameters to end devices is hugely important for the performance of LoRaWAN networks.
Furthermore, the capacity for different types of traffic is studied.
The results confirm recent findings from literature that the limited downstream capacity highly deteriorates the packet delivery ratio of confirmed upstream messages.
Increasing the gateway density can delay the onset of this effect, but it cannot be eliminated completely.
Finally, it is the hope of this work to encourage future work on all aspects of LoRaWAN networks by means of the publicly available \mbox{ns-3} module.
To this end, a number of interesting topics are presented as well.

\section*{Acknowledgment}
This work was carried out in the context of following projects.
MoniCow is a project realized in collaboration with imec. Project partners are DeLaval, Metagam, Multicap, NXP Semiconductors N.V. and snapTonic, with project support from VLAIO (Flanders Innovation \& Entrepreneurship).
IDEAL-IoT (Intelligent DEnse And Longe range IoT networks) is an SBO project funded by the Fund for Scientific Research-Flanders (FWO-V) under grant agreement \#S004017N. 
`Processing visual sensor data in low-power wide area networks' is a project funded by the Fund for Scientific Research-Flanders (FWO-V) under grant agreement \#G084177N.


\ifCLASSOPTIONcaptionsoff
  \newpage
\fi



\bibliographystyle{IEEEtran}
\bibliography{/home/fvdabeele/Documents/library.bib}
%
%
%

\end{document}